# The Influence of Task and Group Disparities over Users' Attitudes Toward Using Large Language Models for Psychotherapy

Qihang He[2], Jiyao Wang[1], Dengbo He[1,3,4]

[1]Thrust of Robotics and Autonomous Systems, Systems Hub, The Hong Kong University of Science and Technology (Guangzhou), Guangzhou, China
[2]Pittsburgh Institute, Sichuan University, Chengdu, China
[3]Thrust of Intelligent Transportation, Systems Hub, The Hong Kong University of Science and Technology (Guangzhou), Guangzhou, China
[4] HKUST Shenzhen-Hong Kong Collaborative Innovation Research Institute, Futian, Shenzhen

The population suffering from mental health disorders has kept increasing in recent years. With the advancements in large language models (LLMs) in diverse fields, LLM-based psychotherapy has also attracted increasingly more attention. However, the factors influencing users' attitudes to LLM-based psychotherapy have rarely been explored. As the first attempt, this paper investigated the influence of task and group disparities on user attitudes toward LLM-based psychotherapy tools. Utilizing the Technology Acceptance Model (TAM) and Automation Acceptance Model (AAM), based on an online survey, we collected and analyzed responses from 222 LLM-based psychotherapy users in mainland China. The results revealed that group disparity (i.e., mental health conditions) can influence users' attitudes toward LLM tools. Further, one of the typical task disparities, i.e., the privacy concern, was not found to have a significant effect on trust and usage intention. These findings can guide the design of future LLM-based psychotherapy services.

## INTRODUCTION

Mental health disorders are becoming more prevalent over the past years. According to the World Health Organization (2023), in 2019, depression affected about 280 million individuals worldwide and 3.94% of the population reported having anxiety disorder. Especially, the population suffering from anxiety and depression has experienced a notable increase due to COVID-19 (World Health Organization, 2022). In the past, some studies attempted to employ traditional natural language models for assisting individuals in monitoring and identifying emotional disturbances (e.g., Emotional Support Conversation (ESC) framework (Liu et al., 2021), EMMA (Li et al., 2021)). Recently, with the fast development of large language models (LLMs), LLM-based online tools have become available for psychotherapy. The superior content generation and comprehension abilities of LLMs compared to traditional natural language models (Ai et al., 2023) proved their great potential to be used in various domains, including code generation (Vaithilingam et al., 2022), academic writing (J. Wang, Huang, et al., 2024; J. Wang, Hu, et al., 2024), and mental health (Kumar et al., 2022). Previous research (Kjell et al., 2023) has found that, in the field of psychotherapy, LLMs presented higher effectiveness in the psychological text analysis and gained deeper insights into individuals' intentions and psychological profiles. Consequently, the utilization of LLMs to help individuals deal with mental health problems has become possible and attracted increasing attention.

However, compared to other tasks (e.g., programming, translation), psychotherapy inherently possesses its distinct characteristics, particularly in aspects such as extra empathy demand and confidentiality concerns (Lustgarten et al., 2020), which we defined as task disparity. Meanwhile, the population with mental disorders exhibits distinctive group-specific characteristics, which we defined as group disparity. Individuals seeking psychotherapy, often suspecting or diagnosed with mental disorders, show distinct attitudes and usage tendencies towards LLMs compared to the general population. For example, when interacting with new technology, anxiety may obstruct rational decision-making, leading to ineffective or even hazardous choices (Peters et al., 2006). Besides, depression and anxiety may have a significant impact on the trust that humans have in automated systems, which could lead to problems with abuse and underutilization (Crawford et al., 2021). Thus, it is necessary to conduct research from the perspective of the attitudes and usage propensities of individuals with mental disorders by exploring the potential influential effects among individual characteristics and psychological variables to enhance the application of LLMs in psychotherapy.

Given that LLMs can be conceptualized as automated systems aiding users in task execution, findings from other human-automation interaction domains may be applied to human-LLM interaction. Thus, in this study, through an online questionnaire in mainland China, we systematically investigated potential typical influential factors of users' trust in and acceptance of LLMs based on some widely used models, including the Technology Acceptance Model (TAM) (Davis, 1989) and Automation Acceptance Model (AAM) (Ghazizadeh et al., 2012). User-related external variables (age, LLMs usage frequency for psychotherapy, education, and personality), users' trust, perceived usefulness, and perceived ease of use towards LLMs in psychotherapy were considered. Additionally, given that the task and group disparity may moderate users' attitudes, we also considered users' privacy concerns and two types of mental disorders as potential moderating in the model. Specifically, following the hierarchy influence pattern in AAM, the influential factors of users' trust were first explored. Then, we investigated how trust towards LLMs in psychotherapy can affect users' intention to use. The findings can guide the design of LLMs in psychotherapy.

Table 1. Descriptive Statistics of Investigated Factors.

| Variable (Abbreviation) | Description | Distribution of extracted variables |
|---|---|---|
| Age | Age of participants. | - Mean: 24.8 years old (SD: 5.7, min: 18, max: 57) |
| Usage Frequency (UF) | How frequency participants use LLMs for psychotherapy. | - Rarely (n=55, 24.7%)<br>- Sometimes (n=94, 42.3%)<br>- Always (n=73, 33.0%) |
| Education (ED) | The education level of participants. | - Below bachelor's degree (n=52, 23.4%)<br>- Bachelor's degree and above (n=170, 76.6%) |
| Anxiety Level (AL) | Seven questions from Hospital Anxiety and Depression Scale (HADS) (Zigmond & Snaith, 1983) for anxiety and depression each. Each question has four choices which scores ranging from 0 to 3.<br>- 0-7: Normal<br>- 8-10: Borderline abnormal (borderline case)<br>- 11-21: Abnormal (case) | - Normal (n=34, 15.3%)<br>- Borderline abnormal (n=96, 43.2%)<br>- Abnormal (n=92, 41.4%) |
| Depression Level (DL) | | - Normal (n=45, 20.3%)<br>- Borderline abnormal (n=136, 61.3%)<br>- Abnormal (n=41, 18.4%) |
| Trust in LLMs (TL) | The sum of the scores of reliability, accuracy, traceability, trust, and dependability (Franke et al., 2015), ranging from 5 to 25. | - Mean: 19.8 (SD: 3.2, min: 11, max: 25) |
| Trust Propensity (TP) | The sum of the scores of five positive statements and one negative statement (Merritt et al., 2019), ranging from 6 to 30. | - Mean: 23.6 (SD: 3.8, min: 12, max: 30) |
| Privacy Concern (PC) | The sum of the scores of four statements (Stewart & Segars, 2002), ranging from 4 to 20. | - Mean: 14.3 (SD: 3.4, min: 4, max: 20) |
| Personality Extraversion (E) | The scores from two opposite questions corresponding to each personality trait were averaged (Gosling et al., 2003), ranging from 1 to 7 for each trait. | - Mean: 4.6 (SD: 1.7, min: 1, max: 7) |
| Personality Agreeableness (A) | | - Mean: 4.7 (SD: 1.0, min: 1.5, max: 7) |
| Personality Conscientiousness (C) | | - Mean: 4.9 (SD: 1.4, min: 1.5, max: 7) |
| Personality Emotional Stability (ES) | | - Mean: 4.5 (SD: 1.4, min: 1.5, max: 7) |
| Personality Openness to Experiences (O) | | - Mean: 4.9 (SD: 1.2, min: 2, max: 7) |
| Perceived Usefulness (PU) | The sum of scores of four statements (Venkatesh & Davis, 2000), ranging from 4 to 20. | - Mean: 16.2 (SD: 2.8, min: 4, max: 20) |
| Perceived Ease of Use (EoU) | The sum of scores of four statements (Venkatesh & Davis, 2000), ranging from 4 to 20. | - Mean: 15.2 (SD: 2.6, min: 6, max: 20) |
| Intention to Use (IU) | The sum of scores of six positive statements and one negative statement (Adell, 2010), ranging from 4 to 20. | - Mean: 15.8 (SD: 3.1, min: 5, max: 20) |

## APPROACH

### Questionnaire Design

Table 1 provides an illustration of the questionnaire design. This questionnaire was designed based on the TAM and AAM while questions were divided into five parts. Part 1 collected user-related external variables (age, LLMs usage frequency for psychotherapy, education, and personality) as they may affect users' attitudes. In part 2, we introduced Hospital Anxiety and Depression Scale (HADS) to regard the states of mental disorders as the group disparity. In part 3, we defined task disparity to reveal users' concerns about privacy issues. Part 4 assessed several significant candidate influential factors chosen from the AAM, including trust, trust propensity, perceived usefulness, and perceived ease of use. Finally, part 5 defined intention to use as the dependent variable. Since LLM users in mainland China with mental disorders were the study's target audience, a Chinese version of the questionnaire was utilized.

### Participants

An online survey study was conducted, and participants were recruited in online forums. In total, 567 participants completed the questionnaire. We eliminated the answers from those who never used LLMs for psychotherapy and screened the answers based on two quality-checking questions (i.e., "If you are answering the question carefully, please select the second/third option"). Finally, 222 participants (144 male and 78 female) with an average age of 24.8 years old (min: 18, max: 57, standard deviation: 5.7) were kept for analysis. They were paid 5 RMB each for their time in completing this 10-minute questionnaire.

### Statistics Models

To account for the hierarchical relationship between trust and intention to use and influence pattern among other variables in AAM, following the hierarchy influence pattern in AAM, two mixed linear regression models were built to successively investigate: 1) influential factors of users' trust in LLMs for psychotherapy (i.e., Trust Model); 2) influential factors of users' intention to use LLMs for psychotherapy (i.e., Intention Model). For the Intention Model, factors that were not significant in the Trust Model were used as predictors, while significant factors in the Trust Model were abandoned.

Both models were built with Proc Mixed in "SAS OnDemand for Academics". All variables (excluding the dependent variables) and their two-way interactions, were utilized as independent variables in initial full models. To prevent multicollinearity issues, we conducted the correlation analysis using the Spearman Correlation among all factors before modeling. Highly correlated (>0.5 or <-0.5) independent variables were aggregated or abandoned, based on the Bayesian Information Criterion (BIC) of the models. Thus, the variables in the full Trust Model included following variables: *Openness to Experiences*, *Education*, *Usage Frequency*, *Trust Propensity*, *Anxiety Level*, and *Depression Level*; and the variables in the full Intention Model were *Anxiety Level*, *Perceived Usefulness*, *Perceived Ease of Use*, and *Trust in LLMs*. Then, a backward stepwise selection was performed based on BIC to obtain the best-fitted models. Post-hoc comparisons were conducted for significant ($p < .05$) main effects and two-way interactions.

### RESULTS

All significant effects in the fitted models after model selection and significant post-hoc contrasts are reported.

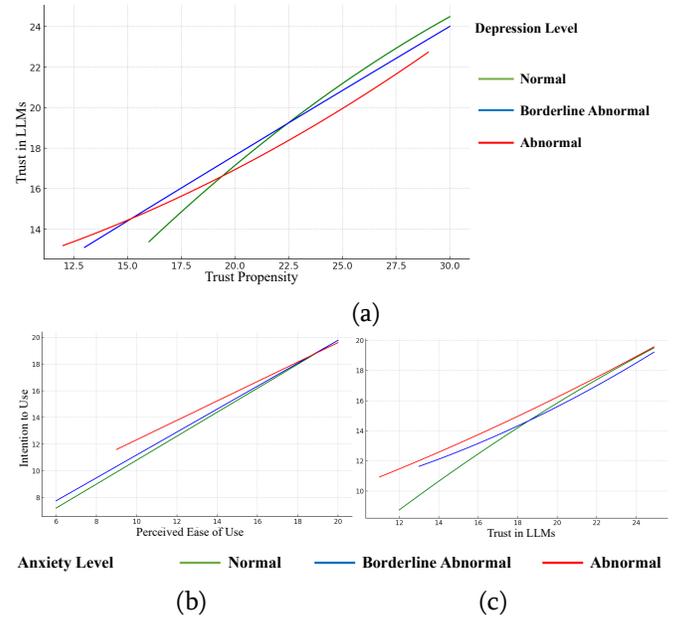

Figure 1. Visualization of significant interaction effects in Trust and Intention Model.

### Trust Model

Table 2. Summary of Trust Model Results

| Independent Variable | F-value | p-value |
|---|---|---|
| Anxiety Level | F(2, 212) = 0.69 | .5 |
| Perceived Usefulness | F(1, 212) = 93.68 | <.0001 ** |
| Perceived Ease of Use | F(1, 212) = 29.26 | <.0001 ** |
| Trust in LLMs | F(1, 212) = 14.17 | .0002 ** |
| Perceived Ease of Use * Anxiety Level | F(2, 212) = 7.13 | .001 ** |
| Trust in LLMs * Anxiety Level | F(2, 212) = 6.73 | .002 ** |

Table 2 presents the results of the Trust Model. Specifically, higher *Openness to Experiences* contributes to higher *Trust in LLMs* with a difference (Δ) of 0.37 and a 95% confidence interval (CI) of [0.11, 0.63], F(1, 206)=8.12, $p$=.005. Besides, lower Usage Frequency of LLMs for psychotherapy led to lower trust in LLMs (Δ=-1.07, 95% CI: [-1.99, -1.34], t(206) = -2.71, $p$=.02).

Figure 1a visualizes the significant interaction effect between *Trust Propensity* and *Depression Level*. Specifically, the positive correlation between *Trust Propensity* and *Trust in LLMs* was weaker among participants with higher levels of depression compared to those with milder symptoms of depression.

### Intention Model

As shown in Table 3, we found that *Perceived Usefulness*, *Perceived Ease of Use*, and *Trust in LLMs* are all significantly associated with *Intention to Use*. Specifically, for each one-unit increase in *Perceived Usefulness*, each one-unit increase in *Perceived Ease of Use*, and every 10 units increase in *Trust in LLMs*, 0.54 units (95% CI: [0.43, 0.65], $F(1, 212) = 93.68$, $p=<.0001$), 0.22 units (95% CI: [0.06, 0.37], $F(1, 212) = 29.26$, $p=<.0001$) and 2.7 units (95%CI: [0.16, 0.38], $F(1, 212) = 14.17$, $p=.0002$) in *Intention to Use* have been observed, respectively.

Table 3. Summary of Intention Model Results

| Independent Variable | F-value | p-value |
|---|---|---|
| Openness to Experiences | F(1, 206) = 8.12 | .005 ** |
| Education | F(1, 206) = 1.80 | .2 |
| Usage Frequency | F(2, 206) = 3.80 | .02 ** |
| Trust Propensity | F(1, 206) = 137.1 | <.0001 ** |
| Anxiety Level | F(2, 206) = 0.82 | .4 |
| Depression Level | F(2, 206) = 2.61 | .08 * |
| Usage Frequency * Depression Level | F(4, 206) = 1.52 | .2 |
| Trust Propensity * Depression Level | F(2, 206) = 3.07 | .049 ** |

Moreover, as shown in Figure 1b, a significant interaction between *Perceived Ease of Use* and *Anxiety Level* was observed for *Intention to Use*. Specifically, With the decrease in *Anxiety Level*, their *Intention of Use* became increasingly positively correlated to their *Perceived Ease of Use*. Further, as shown in Figure 1c, compared to those with a higher *Anxiety Level*, users with a relatively lower *Anxiety Level* were less likely to use LLMs. However, with the increase of *Trust in LLMs*, *Intention of Use* was most positively correlated to the trust in LLMs among people with a normal level of anxiety.

DISCUSSION

Through an online questionnaire in mainland China, based on TAM and AAM models, this paper investigated potential influential factors of users' trust in and acceptance of LLMs for psychotherapy. We also considered how group and task disparities can moderate the effect of the influential factors.

From Trust Model, we found that those who had more experience of using LLMs for psychotherapy and those who presented a higher trust propensity were more likely to trust LLMs. In addition, we found that people who had higher *Openness to Experience* showed a higher trust level. These findings are in line with previous studies targeting users' attitudes toward other technologies (Cai et al., 2022; J. Wang, Tu, et al., 2024). Additionally, from the Trust Model, we surprisingly identified the significant interaction effect between trust propensity and the individual's depression level on users' trust. Specifically, high depression level can eliminate the effectiveness of trust propensity. It is not difficult to understand: when one is experiencing severe depression, they may lose their enthusiasm in the social environment in which they live (Z. Wang & Zou, 2022), and this can influence their attitudes towards things around them, including LLMs. Thus, on the one hand, our observations reveal that individual heterogeneity, including mental health conditions should be considered when understanding one's attitudes towards technologies. On the other hand, it indicates that more efforts should be made for individuals with higher depression levels to trust LLMs.

Then, based on the Intention Model, we validated the applicability of TAM and AAM in explaining users' intention to use LLMs for psychotherapy. As expected, we found that users' intentions can be directly moderated by the evaluation of (i.e., perceived usefulness and perceived ease of use) and trust towards the LLMs. These positive associations are in line with previous research focusing on users' acceptance of technology in general (Hickerson & Lee, 2022). Moreover, as a major contribution of this study, we observed that users' anxiety levels and attitudes towards LLMs can mediate intentions to use the LLM for psychotherapy. In general, people with high levels of anxiety tended to accept LLMs for psychotherapy. It can be explained by the stronger willingness of people with anxiety to control their health than others (Shapiro Jr et al., 1996). What is more interesting is that, when users' perceived ease of use and trust towards LLMs increase, people with less anxiety present a more obvious increasing intention of use. It suggests that, those with lower anxiety levels prefer to use relatively easy-to-use and trustworthy LLMs. Combining this finding and the association between depression and trust, we can conclude that developers should pay special attention to system design or design adaptive interfaces to better serve those with high anxiety or depression levels.

Finally, the task disparity, i.e., the privacy concern we explored in our study, was not found to have a significant effect on trust and usage intention. We assume that it might be because users may inherently trust well-known technology providers (e.g., OpenAI). Future research with more diverse service providers and with data from different countries to better understand the effect of task disparities on users' attitudes toward psychotherapy-related LLM tools.

CONCLUSIONS

In this work, through a survey study, we validated the applicability of TAM and AAM in explaining users' intention to use LLMs for psychotherapy. Further, we also expanded the model by incorporating the influence of mental health conditions. Finally, we found that individual heterogeneity should be considered when designing the LLM tools for psychotherapy to better serve people with psychological problems, for example, by adopting the adaptive interface design.

## DECLARATION OF CONFLICTING INTEREST

The authors declare that they have no known competing financial interests or personal relationships that could have appeared to influence this paper.

## FUNDING

This work was supported by the Guangzhou Municipal Science and Technology Project (No. 2023A03J0011), and the Project of Hetao Shenzhen-Hong Kong Science and Technology Innovation Cooperation Zone (HZQB-KCZYB-2020083).